\documentclass[conference]{IEEEtran}
\IEEEoverridecommandlockouts
\usepackage{hyperref}
\usepackage{cite}
\usepackage{amsmath,amssymb,amsfonts}
\usepackage{algorithmic}
\usepackage{graphicx}
\usepackage{textcomp}

\usepackage{xcolor}
\usepackage{multirow}
\usepackage{dblfloatfix}  
\def\BibTeX{{\rm B\kern-.05em{\sc i\kern-.025em b}\kern-.08em
    T\kern-.1667em\lower.7ex\hbox{E}\kern-.125emX}}
\begin{document}

\title{Reliable Deep Learning based Localization with CSI Fingerprints and Multiple Base Stations}

\author{\IEEEauthorblockN{Anastasios Foliadis\IEEEauthorrefmark{1}\IEEEauthorrefmark{2}, Mario H. Casta\~{n}eda Garcia\IEEEauthorrefmark{1}, Richard A. Stirling-Gallacher\IEEEauthorrefmark{1},  Reiner S. Thom\"a\IEEEauthorrefmark{2}}
	\IEEEauthorblockA{\IEEEauthorrefmark{1}\textit{Munich Research Center}, \textit{Huawei Technologies Duesseldorf GmbH}, 
		Munich, Germany \\
		\textit{\IEEEauthorrefmark{2}Electronic Measurements and Signal Processing}, \textit{Technische Universit\"at Ilmenau}, Ilmenau, Germany\\
		\{\href{mailto:anastasios.foliadis@huawei.com}{anastasios.foliadis}, 
		\href{mailto:mario.castaneda@huawei.com}{mario.castaneda}, 
		\href{mailto:richard.sg@huawei.com}{richard.sg}\}@huawei.com, 
		\href{mailto:reiner.thomae@tu-ilmenau.de}{reiner.thomae@tu-ilmenau.de}}}

\maketitle

\begin{abstract}
Deep learning (DL) methods have been recently proposed for user equipment (UE) localization in wireless communication networks, based on the channel state information (CSI) between a UE and each base station (BS) in the uplink. With the CSI from the available BSs, UE localization can be performed in different ways. One the one hand, a single neural network (NN) can be trained for the UE localization by considering the CSI from all the available BSs as one overall fingerprint of the user's location. On the other hand, the CSI at each BS can be used to obtain an estimate of the UE's position with a separate NN at each BS, and then the position estimates of all BSs are combined to obtain an overall estimate of the UE position. In this work, we show that UE localization with the latter approach can achieve a higher positioning accuracy. We propose to consider the uncertainty in the UE localization at each
 BS, such that overall UE's position is determined by combining the position estimates of the different BSs based on the uncertainty at each BS. With this approach, a more reliable position estimate can be obtained in case of variations in the channel.
\end{abstract}


\section{Introduction}

Interest in user positioning has grown significantly in the past few years due to the increasing demand of applications requiring accurate location information of the user equipment (UE). The precise location information can additionally aid and improve the performance of the current (5G mobile network) and of future communication networks. 

User positioning is particularly required in dense urban environments and indoors where the Global Navigation Satellite System (GNSS) is not available. In such areas, the use of wireless communication networks like 5G can be considered for positioning. One of the benefits of 5G networks over previous standards  is the support of multiple antennas and mm-Wave frequencies with large available bandwidths, which can be leveraged for positioning.

Furthermore the ability to collect large amounts of data makes Deep Learning (DL) based solutions an attractive possibility. DL based localization in wireless networks has shown promising results achieving a sub-centimeter accuracy in some indoor scenarios  \cite{Foliadis2021CSIBasedLW} \cite{Sobehy2019CSIBI}. The idea behind these approaches is that the channel state information (CSI) between the UE and a base station (BS) is considered a fingerprint of the user’s location. A neural network (NN) can be trained on a database of uplink CSI fingerprints with their respective UE's location label, such that the NN can later be used for mapping the CSI of a UE to the UE’s position. 

In the uplink of a communication network, several BSs may be able to measure the CSI of a UE. With the CSI from the available BSs, the UE localization can be performed with an early fusion or late fusion approach. In early fusion, the combined CSI of all the multiple BSs is considered as a single fingerprint that is used as input of a DL model trained for the UE localization. In late fusion, the CSI of each BS is taken as an input of a separate DL model, to obtain an estimate of the UE position at each BS. Afterwards, the position estimates of all BSs are combined to obtain an overall estimate of the UE position. For DL based applications, there is no conclusive evidence as to which type of fusion is better \cite{Ramachandram2017DeepML}, as this is usually problem dependent.


Although typically DL techniques do not provide any information about the reliability of their estimation, there exist techniques that are able to address this issue. The authors of \cite{Gonultacs2020CSIBasedMA} use a late fusion approach with a probabilistic description of the NN's output which contains information about the reliability of the estimation. An uncertainty quantification technique that is based on variational autoencoders (VAE) is used in \cite{Stahlke2021ETRVA} to determine the reliability of the estimations. Other simpler techniques to compute the uncertainty of the estimation include Monte Carlo Dropout (MCD) \cite{Gal2016DropoutAA} and deep ensembles (DE) \cite{Lakshminarayanan2017SimpleAS}. 
DE are based on the observation that when the estimations of multiple DL models (i.e. multiple NNs) differ significantly then the estimation is unreliable. MCD is based on a similar idea as explained in \cite{Gal2016DropoutAA}.




In this work, we show that UE localization with late fusion achieves a higher positioning accuracy compared to early fusion. We propose to have a separate DL model to estimate the UE position at each BS and to estimate the uncertainty of each DL model with the given CSI. The UE’s position is then obtained by combining the estimates of the different BSs while taking into account the uncertainty at each BS. In general, the BSs with the most reliable estimates have a larger impact on determining the UE's position. We also show that with our proposed approach, a more reliable position estimate can be obtained in case of variations in the channel. We model the channel variations with a random blockage of the line of sight (LOS) path of the channel between the UE and a BS.

The paper is structured as follows. In section \ref{sec:system model} the considered system model is described and the late and early fusion approaches are introduced. In section \ref{sec:uncertainty} the uncertainty estimation methods are explained and in section IV we describe the simulation setup along with the NN structure. The results and conclusion are then presented in sections V and VI respectively. 

\section{System Model}
\label{sec:system model}
\subsection{System Model}
\label{sytem_model}
We consider an uplink setup with $N_B$ BSs each with $N_R$ Rx antennas and a single Tx antenna at the UE. The UE transmits a reference signal on $N_C$ subcarriers within an orthogonal frequency division multiplexing (OFDM) symbol. The transmitted reference signal that is received at a BS, is used to estimate the CSI between the UE and the BS. We assume that each of the $N_B$ BSs can estimate the channel from the UE. The estimated CSI between the $n$-th BS and the UE can be described as:

\begin{figure}[b!]
	\centering
	\includegraphics[scale=0.65]{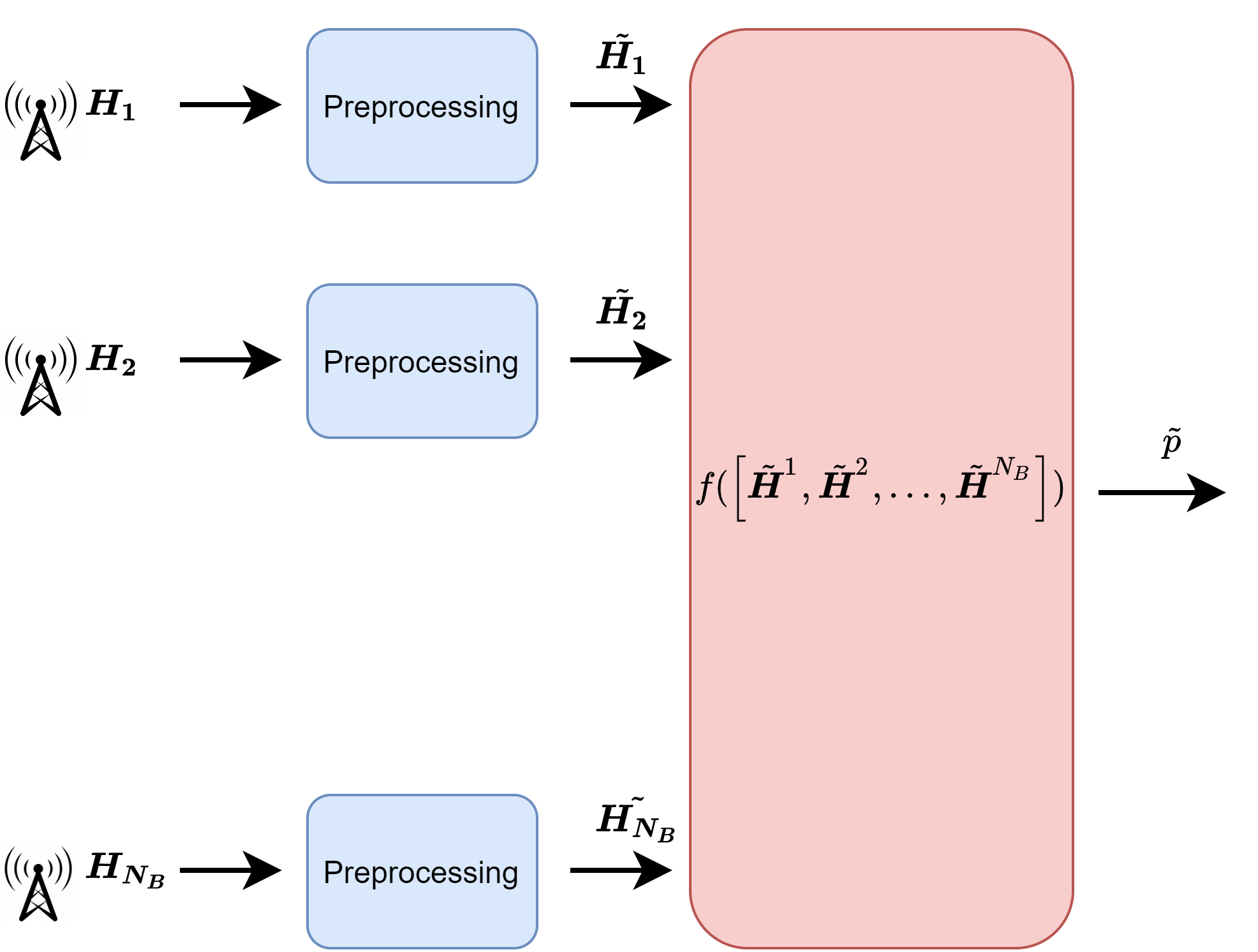}
	\caption{Early Fusion Block Diagram}
	\label{fig:earlyfusion}
\end{figure}

\begin{equation}
	\boldsymbol{H}^n = [\boldsymbol{h}_0^n, \boldsymbol{h}_1^n, ..., \boldsymbol{h}_{N_C - 1}^n] \in \mathbb{C}^{N_C \times N_R},
\end{equation}
where $\boldsymbol{h}^n_m \in \mathbb{C}^{N_R}$ is the vector with the estimated uplink channels between the UE and the $N_R$ antennas at the $n$-th BS and at $m$-th subcarrier. The estimated channel $\boldsymbol{H}^n$ depends on the multipath between transmitter and receiver and it could be considered a distinct fingerprint of the position of the UE. The CSI fingerprints are processed before being inputted to the DL model, mainly due to the fact that they need to be represented by real values at the input of the NN. We denote the processed fingerprint input as $\tilde{\boldsymbol{H}}^n$

\subsection{DL Based Positioning with Fingerprints}

Deep learning based localization using fingerprint inputs of the environment is generally divided into two phases. During the first phase, referred to as training phase, the measurements $\boldsymbol{H}$ that constitute the fingerprints are gathered and stored into a database along with their respective $D$-dimensional position label $\boldsymbol{p} \in \mathbb{R}^D$. The data are then used as training inputs to a DL model which learns to map the measurement to the position label. During the second phase, the testing phase, every subsequent CSI measurement is passed onto the DL model, which predicts the location. The DL model $f(\boldsymbol{H})=\tilde{\boldsymbol{p}}$ is trained to minimize the mean squared error (MSE) between the position label and the estimated position $\tilde{\boldsymbol{p}}$:

\begin{equation}
	L(\theta) = \frac{1}{D}\sum_{d=1}^{D} |\boldsymbol{p}_{d} - \tilde{\boldsymbol{p}}_{d}|^2
\end{equation}
with respect to the DL model parameters $\theta$.


\subsection{Late and Early Fusion}

\begin{figure}[b!]
	\centering
	\includegraphics[scale=0.6]{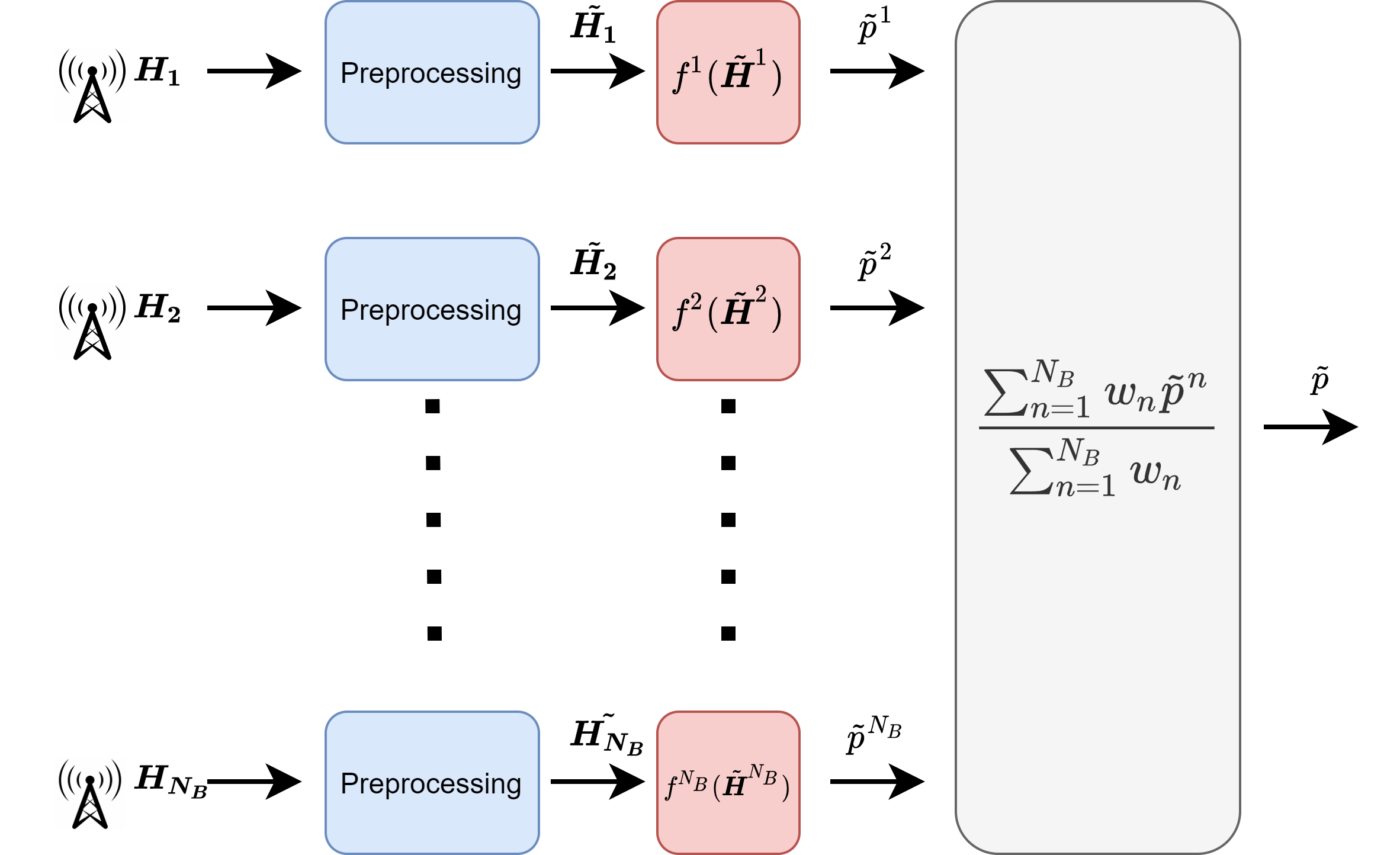}
	\caption{Late Fusion Block Diagram}
	\label{fig:latefusion}
\end{figure}
With the CSI of a UE available at multiple BSs, there is the choice between late and early fusion, for performing the UE localization. In early fusion, the preprocessed CSI matrices of
all the $N_B$ BSs are concatenated into a single fingerprint $[\tilde{\boldsymbol{H}}^1, \tilde{\boldsymbol{H}}^2, ..., \tilde{\boldsymbol{H}}^{N_B}]$ that is used as input of a single DL model $f([\tilde{\boldsymbol{H}}^1, \tilde{\boldsymbol{H}}^2, ..., \tilde{\boldsymbol{H}}^{N_B}])$. Based on a database of such fingerprints for different UE positions, the DL model is trained to determine the UE's position. Thus, we have that $f([\tilde{\boldsymbol{H}}^1, \tilde{\boldsymbol{H}}^2, ..., \tilde{\boldsymbol{H}}^{N_B}]) = \tilde{\boldsymbol{p}} \in \mathbb{R}^D$. A block diagram of early fusion is shown on Fig. \ref{fig:earlyfusion}.

For late fusion, the preprocessed CSI matrix $\tilde{\boldsymbol{H}}^n$ of the $n$-th BS is taken as input of a DL
model $f^n(\tilde{\boldsymbol{H}}^n)$ and is used to obtain an estimate $\tilde{\boldsymbol{p}}^n$ of the UE position ${\boldsymbol{p}}$ at each
BS, i.e. $f^n(\tilde{\boldsymbol{H}}^n)=\tilde{\boldsymbol{p}}^n$.  The DL model at each BS is trained based on a database of preprocessed CSI from that BS for different UE positions. Afterwards, the position estimates of all BSs are combined
to obtain an estimate of the UE position ${\boldsymbol{p}}^n$. A simple visualization of the late fusion approach is given in Fig. \ref{fig:latefusion}. Since the combination of the predictions from multiple models is performed at the decision stage, this type of data
fusion is called late fusion.


 The most straight-forward way to combine the $N_B$ different estimates is using a weighted average for each element $d \in [0, ..., D-1]$ of the output vector with weights $W_d = [w_{d,1}, w_{d,2}, ..., w_{d,N}]$ resulting in:

\begin{equation}
	\tilde{{p}_d} = \frac{\sum_{n=1}^{N_B} w_{d, n} \tilde{{p}_d}^n}{\sum_{n=1}^{N_B} w_{d,n}
	}.
	\label{eq:weighted_average}
\end{equation}

When $w_{d, 1} = w_{d, 2} = ... = w_{d, N}$, this results in an ordinary average of the position estimates. Even though this method has shown promising results in some occasions \cite{Sobehy2019CSIBI}, it may not be robust to environmental changes. To create a more robust model we propose to determine the weights based on the uncertainty information. The least certain models should be weighted less and vice-versa.


It should be noted here that from implementation perspective, late fusion has the advantage that it can be implemented both as a distributed scheme as well as centrally processed scheme. The disadvantage of a centrally processed scheme is that a large amount of data has to be transmitted from the BSs to the central location server. This downside is always present in the early fusion approach and hence one may have to resort to compression techniques if this approach is used. 
\section{Uncertainty Estimation}
\label{sec:uncertainty}

For UE localization based on CSI fingerprints, it is expected that for a UE at a given position, the CSI measured during the training phase and the testing phase are similar. However, the CSI between the two phases can differ considerably due to variations in the wireless channel. Movement of objects in the environment can cause a scatterer to change position or a path to be blocked, e.g. the LOS path can be blocked.


 We propose to address this issue by using the late fusion approach, i.e. by using a separate DL model at each BS. Furthermore, by adding uncertainty quantification techniques at each model, the uncertainty in the UE localization with the CSI at that BS can be estimated. If there is a change in the CSI for a given UE position at a BS (compared to the training phase), the proposed approach could report a low certainty. By applying such approach at each BS, the uncertainty estimation will provide information about the certainty that each BS has about its position estimate. 

Uncertainty estimation with early fusion is also possible to determine whether an estimate of the UE's position is reliable or not. This is in contrast to uncertainty estimation with late fusion where an estimate may be still be possible even if the estimations from some BSs are unreliable.

\subsection{Aleatoric Uncertainty}

There are two main sources of uncertainty, which are called epistemic and aleatoric uncertainty \cite{Kendall2017WhatUD}. Aleatoric uncertainty refers to the uncertainty which is inherent in the data such as noise or additionally in the case of DL-based UE localization due to indistinguishable fingerprints.

Aleatoric uncertainty can be further separated into two kinds: homoscedastic aleatoric uncertainty and heteroscedastic aleatoric uncertainty. The former describes uncertainty which is constant regardless of the input to the NN, i.e. every input exhibits the same amount of noise. Heteroscedastic uncertainty on the other hand is input dependent, meaning that some inputs contain less information about the output than others. In the context of DL localization this could be due to variable path loss, which depends on the distance between UE and BS, or on insufficient information from the multipath.  

For the task of localization using CSI fingeprints, the noise affects each input differently, therefore in this scenario the heteroscedastic aleatoric uncertainty needs to be considered. Since aleatoric uncertainty is data dependent the loss function $L(\theta)$ has to be modified to include the uncertainty of the $d$-th output component in the position estimate. The uncertainty is given by $s_d = \log(\sigma_d^2)$, where $\sigma_d^2$ is the variance of $d$-th component of the position estimate. In this way, the NN learns which data provide less information \cite{Kendall2017WhatUD}. The modified loss function is given by:

\begin{equation}
 L'(\theta) = \frac{1}{2D}\sum_{d=1}^{D} \exp(-s_{d})|\boldsymbol{p}_{d} - \tilde{\boldsymbol{p}}_{d}|^2 + s_d.
 \label{eq:aleatoric_loss}
 \end{equation}
The loss function consists of two components. The first term is a regression term which increases the uncertainty $s_d$ when the MSE becomes larger, i.e. when there is insufficient information at the input sample,  and the last term is a regularization term which limits an infinite increase of the loss function. Subsequently the model output is expanded to include the uncertainty $\boldsymbol{s} \in \mathbb{R}^D$ as well, i.e.  $f(\tilde{\boldsymbol{H}})=[\tilde{\boldsymbol{p}},\boldsymbol{s}]$, where $\boldsymbol{s} = [s_1, s_2, ..., s_D]$.

\subsection{Epistemic Uncertainty}
\label{sec:epistemic uncertainty}
Epistemic uncertainty refers to the uncertainty that the model experiences due to insufficient training data. For example a DL localization model would exhibit high epistemic uncertainty for a region in space where no fingerprints where collected. It can be useful to identify some change in the measured CSI, which would indicate a change in the environment.

\subsubsection{Monte Carlo Dropout (MCD)}
One way to capture the epistemic uncertainty of the model is the MCD technique. It is a practical approach to obtain Bayesian approximation of the model posterior distribution. When using dropout, individual neurons of every weight layer of the DL model are removed with probability $p$. Normally, dropout is applied only during training but with MC dropout, the same dropout probability is retained also during testing. Effectively each forward pass through the NN results in a new configuration (since different neurons are dropped) and performing multiple forward passes is equivalent to sampling from the approximate posterior distribution of the model parameters given the dataset \cite{Gal2016DropoutAA}. The variance of the estimates from the different NN configurations of each of the forward passes is a measure of the uncertainty. 

Finally, after $T$ forward passes, the combined aleatoric and epistemic uncertainty for the $d$-th component of the position estimate of the $n$-th model can be calculated as \cite{Kendall2017WhatUD}:

\begin{equation}
	(\sigma_{MC, d}^n)^2 \approx \frac{1}{T}\sum_{t=1}^{T}\left(\tilde{\boldsymbol{p}}_{t,d}^n\right)^2 - \left(\frac{1}{T}\sum_{t=1}^{T}\tilde{\boldsymbol{p}}_{t,d}^n \right ) ^2 + \frac{1}{T} \sum_{t=1}^{T} \exp(s_{t, d}^n)
	\label{eq:predicted_variance}
\end{equation}
where $t$ indicates the current forward pass. The last term represents the aleatoric uncertainty, while the first two terms are used to compute the variance of the estimates from the $T$ forward passes. Having this measure of uncertainty, the weighted average for late fusion can by calculated by setting $w_{d, n} = 1/(\sigma_{MC, d}^n)^2$ in \eqref{eq:weighted_average}.
\subsubsection{Deep Ensemble (DE)}

One other approach to estimate uncertainty is the use of DE. An ensemble of NNs is a set of models which are trained for the same task. The individual models are then combined to achieve a better estimation. They consistently outperform single NNs particularly when there is diversity between the models \cite{Sobehy2019CSIBI}. In a DL setting, diversity can be achieved by the randomly initialized weights of the NN, by modifying the hyperparameters of the algorithm or by varying the training data. 

For UE localization with multiple BSs  each BS observes a different CSI for a UE position and hence, the training data available at each BS is different. Thus, in the case of late fusion, we obtain different models simply by training them on the different training data sets from the $N_B$ BSs. The different models would form an ensemble of NNs, whose outputs are combined to provide an estimate of the UE position.

With the $N_B$ trained models, we propose to group the BSs into $N_B$ sets each with $N_B-1$ BSs, i.e. in each set one BS is excluded. We define the $n$-th set as $S_n = \{i \quad | i \in [1, ..., N_B] \backslash \{n\}\}$. We can obtain a different ensemble of $N_B -1$ NNs by considering the CSI from the BSs within each set $S_n$. For each of those sets, and by including the aleatoric uncertainty, the variance of the $N_B-1$ position estimates from the BSs within the set is computed as \cite{Lakshminarayanan2017SimpleAS}:
\begin{equation}
	(\sigma_{E, d}^n)^2 = \sum_{i \in S_n} \left(\tilde{\boldsymbol{p}_d}^i\right)^2 - \left(\sum_{i \in S_n} \tilde{\boldsymbol{p}_d}^i\right)^2 + \sum_{i \in S_n}\left(\sigma_d^i\right)^2.
\end{equation}
This variance can be thought of as the uncertainty of that particular set which excludes the estimation of the model $n$. For this reason we consider the variance $(\sigma_{E, d}^n)^2$ divided by the sum of the variances of all other $N_B-1$ sets as a measure of the certainty of the model $n$. Thus, for the DE approach we set the weights of \eqref{eq:weighted_average} of the late fusion equal to 
\begin{equation}
	w_{d, n} = \frac{(\sigma_{E, d}^n)^2}{\sum\limits_{i \in [1, ..., N_B] \backslash \{n\}}(\sigma_{E, d}^i)^2}
	\label{eq:DE weight}
\end{equation}
The reasoning behind this measure, is that if the uncertainty of one set is high compared to all other sets then this means that the excluded BS from that set is the most certain. The opposite is also true. By setting the weight as in eq. \eqref{eq:DE weight} then we are determining the most certain BSs, by essentially comparing the uncertainties of the sets.

\section{Simulation Setup}

\subsection{Database Description}

\label{sec:database description}
To compare the early fusion and the late fusion approaches, along with the proposed uncertainty estimation, we used ray-tracing dataset called DeepMIMO \cite{Alkhateeb2019DeepMIMOAG}. With the DeepMIMO dataset, the user can define different channel parameters and simulate a propagation scenario from a UE at a given location to several BSs in an outdoor scenario. The channel including the path loss is then generated using accurate ray-tracing data. 

\begin{figure}[t!]
	\centering
	\includegraphics[scale=0.5]{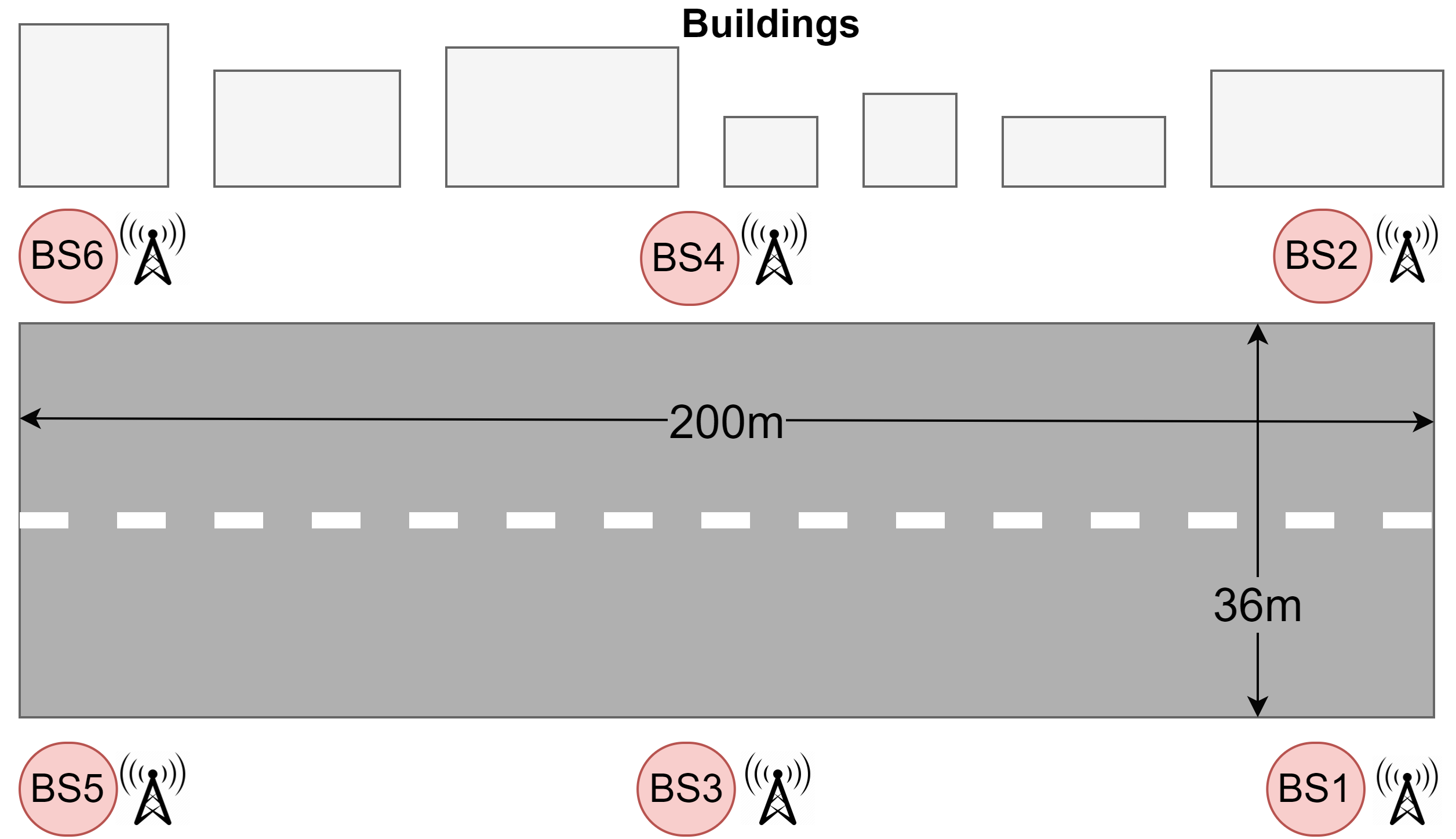}
	\caption{Early Fusion Block Diagram}
	\label{fig:deepmimo}
\end{figure}

For our evaluations, we consider only one segement of a street within the urban outdoor scenario from the DeepMIMO dataset. Namely, we consider the street with $N_B=6$ BSs as depicted in Fig. \ref{fig:deepmimo}. Each BS has a uniform linear array (ULA) in the azimuth plane with $N_R=16$ antennas and a half wavelength spacing between adjacent antenna elements. The area of the considered scenario is $200 \times 36\text{m}^2$. The carrier frequency of this scenario is $f_c = 3.5\text{GHz}$ and we assume that the bandwidth was set at $BW=80 \text{MHz}$. The number of subcarriers was set at $1024$. We assume the reference signal is transmitted on every 10th subcarrier, whichresults in a CSI fingerprint of $N_C = 103$ subcarriers. We also assume that the UE's transmit power is 23 dBm, the noise floor is -174 dBm/Hz and the receiver's noise figure is 2 dB. 

As already mentioned, in a real environment, the CSI fingerprint at a given UE location may change due to variations in the wireless channel, i.e. movement of objects or scatterers, or blockage of a path due to obstacles. We model the changes in the channel by assuming that the LOS path in the channel between a UE and a BS can be blocked with a given probability. The blockage probability of the LOS path usually increases as the distance between the BS and UE increases, as discussed in \cite{38.901}. Based on this, we assume that the LOS is blocked according to the LOS blockage probability in the warehouse scenario in \cite{38.901}, which is given by:

\begin{equation}
	P_{NLOS} = 1 - \exp \left (- \frac{r \log(1 - 0.05)}{10} \right ),
	\label{eq:nlos_prob}
\end{equation}
where $r$ is the distance between the UE and the BS.

Even though the simulated ray-tracing scenario is an outdoor setting, we believe that the blockage probability for the LOS path in the warehouse scenario as described by \cite{38.901} better captures our intended simulation environment, i.e. people or objects causing signal blockage.

\subsection{Neural Network Setup}

\begin{figure}[t!]
	\centering
	\includegraphics[scale=0.6]{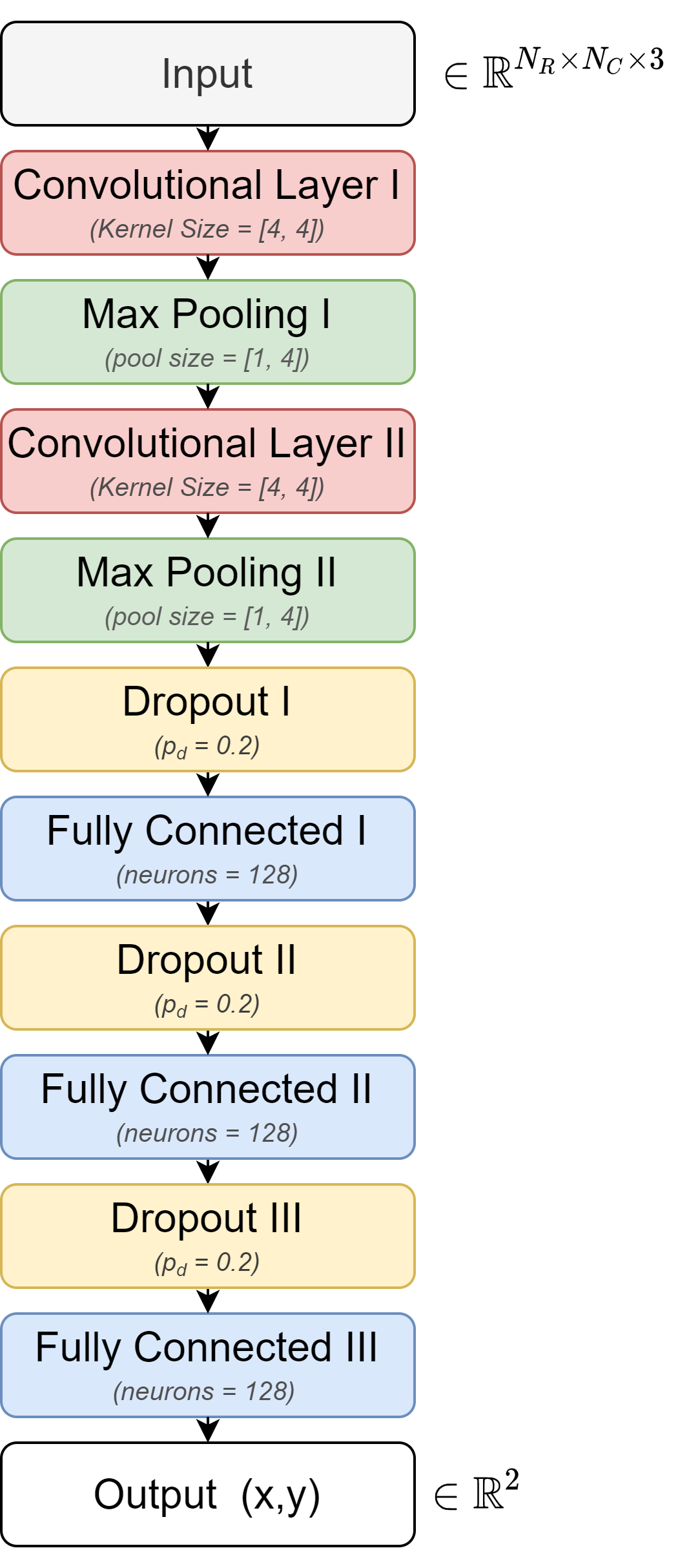}
	\caption{Neural Network Model}
	\label{fig:cnn}
\end{figure}

We consider a convolutional neural network (CNN) with two convolutional layers and two max pooling layers followed by three dense layers as shown in Fig. \ref{fig:cnn} . The input of the CNN, which is the measured CSI $\boldsymbol{H}^{n} \in \mathbb{C}^{N_R\times N_C}$, is processed using the phase difference between the antennas as described in \cite{Foliadis2021CSIBasedLW}, to resolve any timing errors between UE and each BS. This results in a real $3$-dimensional matrix, i.e $\tilde{\boldsymbol{H}}^{n} \in \mathbb{R}^{N_R\times N_C \times 3}$, where the third dimension consists of the magnitude and the sine and cosine of the phase difference between adjacent antennas as separate stacked 2D matrices.  
The two convolutional layers of the NN contain 32 different kernels with dimensions $4 \times 4$ and the dimension of the max pooling layers is $4 \times 1$, only pooling the subcarrier dimension. The outputs of the second pooling layer are vectorized and inputted into 3 dense layers of 128 neurons each with the rectified linear unit as their activation function. A last dense layer with 2 neurons is at the output, with no activation function. Before every dense layer there is a dropout layer with dropout probability $p_d = 0.2$. The same NN structure is considered for both early and late fusion by simply changing the input dimension to correspond to the considered fingerprint. 

The NNs were trained with a batch size of 32 and 64, each for 1000 epochs. The loss function used for training is the one described in Eq. \eqref{eq:aleatoric_loss}. The training set was $80\%$ of the database, while the other $20\%$ was used for the testing. All the input data are normalized in the range $[0,1]$. For the evaluation of the different schemes, we consider the mean error (ME) of the UE positioning, given by the Euclidean distance between the estimated postion and the true position of a UE in the test set.

\section{Simulation Results}

Based on the setup described in Sec. \ref{sec:database description} and Fig. \ref{fig:deepmimo}, we consider two scenarios for the simulation results: a static scenario and a dynamic scenario. For the static scenario, there is no change in the channel between the training phase and the test phase, while for the dynamic scenario there can be a change in the channel between the two phases. For the training phase in both scenarios and also for the testing phase in the static scenario, CSI samples for UE positions are taken within the area depicted in Fig. \ref{fig:deepmimo}, with a LOS and multipath between each UE position and each BS according to the DeepMIMO dataset. However, for the testing phase in the dynamic scenario, the CSI samples are taken assuming that the LOS path to a BS can be blocked according to the probability given in \eqref{eq:nlos_prob}. If this is the case, the LOS path is simply removed from the channel generated from the DeepMIMO dataset.
	\begin{table}[b]
	\setlength\tabcolsep{4pt}
	\caption{Performance with Separate NNs}
	\centering
	\begin{tabular}{||c c c c c c c||} 
		
		\hline
		& \textbf{BS1} & \textbf{BS2} & \textbf{BS3} & \textbf{BS4} & \textbf{BS5} & \textbf{BS6}  \\ [0.5ex] 
		\hline\hline
		\textbf{Static ME (m)} & 2.9227 & 2.8422 & 2.6561 & 2.3191 & 2.5918 & 2.6686 \\ 
		\hline
		\textbf{Dynamic ME (m)} & 5.5495 & 5.9869 & 3.9249 & 3.7786 & 4.7347 & 4.9414 \\ [1ex] 
		\hline
		
	\end{tabular}
	\label{single_bs}
\end{table}
\begin{table}[b]
	\setlength\tabcolsep{6pt}
	\caption{Performance with Data Fusion}
	\centering
	\begin{tabular}{||c c c c c||} 
		\hline
		&\multirow{2}{*}{\textbf{Early Fusion}}  & \multicolumn{3}{ c ||}{\textbf{Late Fusion}}\\
		&   & {Average} & {MCD}  & {DE}\\ [0.5ex] 
		\hline\hline
		\textbf{Static ME (m)} &  1.7532 & 1.4137 & 1.2431 & 1.2777  \\ 
		\hline
		\textbf{Dynamic ME (m)} &  2.5116 & 3.0578 & 1.9864  & 1.9262 \\ [1ex] 
		\hline
		
	\end{tabular}
	\label{data_fusion}
\end{table} 

 We see the results for the NNs trained on each of the 6 BSs in Table \ref{single_bs}. As expected the performance of the UE localization at each BS deteriorates when going from the static scenario to the dynamic scenario, as the CSI in test data may differ from the CSI in the training data. The variations in the performance of the separate NNs for each BSs, may result from the different building layout along the street.

On Table \ref{data_fusion} we see the results for the different data fusion approaches. First it is interesting to see that in a static environment, the early fusion is outperformed by the late fusion approaches, including the simple approach with equal weighting. For late fusion, we consider the combining in \eqref{eq:weighted_average} with different  weighting, namely: equal weighting (averaging) and weighting based on the MCD and DE approaches discussed in Sec. \ref{sec:epistemic uncertainty}. For the static scenario, it is more beneficial to combine different position estimates from the BSs, instead of considering one larger CSI fingerprint for the UE localization. Among the late fusion approaches, better performance is observed with weighting based on uncertainty estimation, even though there are no changes in the channel.

On the other hand, when considering the dynamic scenario, early fusion outperforms late fusion with simple averaging, which implies that the former is more robust to changes in the channel. In this case, the larger CSI fingerprint in early fusion is beneficial, as a channel change at one or two BSs only impacts portions of the large CSI fingerprint, which may allow the NN to still obtain an estimate of the UE position. In case of late fusion with simple averaging, a channel change at one BS may lead to a very poor estimate of the UE position from that BS. Such poor estimate can then dominate the average when combining the estimates from other BSs. 
 
However, the situation changes when using weights based on uncertainty estimation. With MCD or DE weighting,  late fusion outperforms the early fusion in the dynamic scenario. With uncertainty estimation, the BSs that have experienced a channel change have a smaller weight. Hence, their estimate of the UE position would have a reduced impact for the combination when determining the overall UE's position.

Both in the static and the dynamic environment the error becomes smaller when also considering the uncertainty of each model as described in section \ref{sec:epistemic uncertainty}. This makes sense as with uncertainty estimation it is possible to identify which of the models experiences some environmental change. MCD weighting shows slightly better performance in a static environment while the opposite is true for DE weighting. 

\begin{figure}[tb]
	\centering
	\includegraphics[scale=0.5]{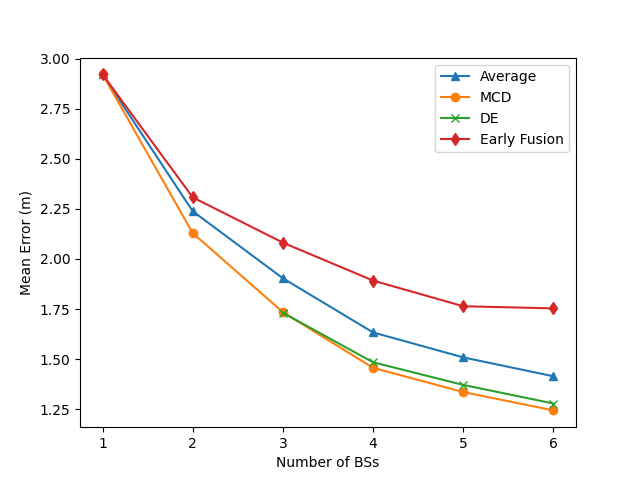}
	\caption{Error of data fusion techniques on static scenario}
	\label{fig:noblockage_error_bs}
\end{figure}

\begin{figure}[tb]
	\centering
	\includegraphics[scale=0.5]{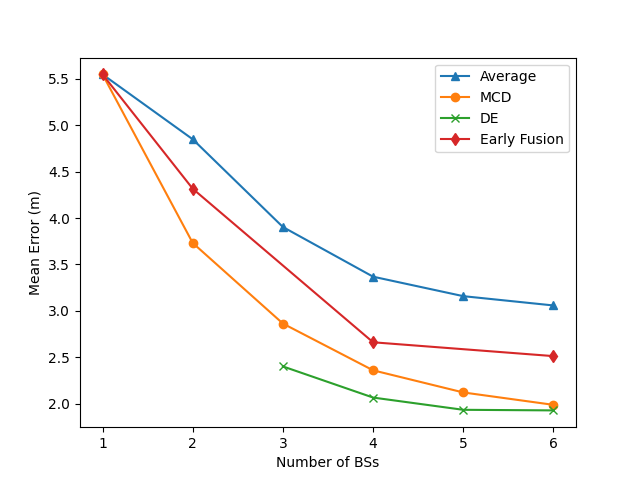}
	\caption{Error of data fusion techniques on dynamic scenario}
	\label{fig:blockage_error_bs}
\end{figure}

We finally evaluate the impact of the number of BSs on the different fusion approaches for the static and dynamic scenario in Fig. \ref{fig:noblockage_error_bs} and \ref{fig:blockage_error_bs}, respectively. The BSs that are considered are selected based on increasing indices, i.e. for 3 BSs, BS 1, BS 2 and BS 3 from Fig. 3 are selected. Based on the description in Sec. III.B, late fusion with DE weighting needs at least 3 BSs to compute the certainty. For only one BS, early fusion, simply averaging and using MCD weighting is exactly the same, although with only 2 BSs the uncertainty estimation in the MCD weighting is already beneficial for both static and dynamic scenarios. In the static scenario, MCD weighting is slightly better than DE weighting for different number of BSs. In the dynamic scenario, however, DE weighting outperforms MCD weighting, with a larger gap for smaller number of BSs. This indicates that the DE weighting is better at identifying the epistemic uncertainty which is higher in the dynamic scenario. Lastly, we see that early fusion is more robust to variations in the channel compared to late fusion with simple averaging, regardless of the number of BSs.



\section{Conclusion}

We compared UE localization with multiple BSs using early fusion and late fusion, and showed that the late fusion achieves a higher positioning accuracy. We propose to have a separate DL model to estimate the UE position at each BS and to estimate the uncertainty of each DL model with the given CSI. Afterwards, the UE's position is determined by combining the estimates of the different BS while taking into account the uncertainty at each BS. Simulations results confirm that our proposed approach is able to deliver a more reliable position estimate even when considering variations in the channel.

%

\bibliographystyle{IEEEtran}
\bibliography{references}

\end{document}